\documentstyle[epsfig,a4,11pt]{article}

\def \lta {\mathrel{\vcenter{\hbox{$<$}\nointerlineskip\hbox{$\sim$}}}}
\def \gta {\mathrel{\vcenter{\hbox{$>$}\nointerlineskip\hbox{$\sim$}}}}

\normalsize

\begin{document}

\title{\bf On the Construction of \\ Quintessential Inflation Models}

\author{Marco Peloso $\! ^*$ and Francesca  Rosati $\! ^\dagger$ }
\maketitle

\begin{center}
\footnotesize{\sl SISSA, via Beirut 2-4, I-34013 Trieste, Italy \\ 
INFN, Sezione di Trieste, Padriciano 99, I-34012 Trieste, Italy} 
\vspace{.15in} \\
\small{ $^* \!$ e-mail: peloso@sissa.it ,
$\;\;\;\; ^\dagger \!$ e-mail: rosati@sissa.it}  
\end{center}

\abstract{Attention has been recently drawn towards models in which
inflation and quintessence schemes are unified. In such
`quintessential inflation' models, a unique scalar field
is required to play both the role of the inflaton and of the
late--time dynamical cosmological constant.
We address the issue of the initial conditions for quintessence in
this context and find that, in the two explicit examples provided,
inflation can uniquely fix them to be in the allowed range for a
present day tracking.}

\vskip0.1in
\noindent
SISSA/99/EP/93

\vskip0.3in

\section{Introduction}

The very last years have witnessed growing interest in cosmological models 
with $\Omega_m \sim 1/3$ and $\Omega_\Lambda \sim 2/3$, following
the most recent observational data (see for example \cite{obs} and 
references therein). 
A very promising candidate for a dynamical cosmological constant 
$\Lambda(t)$ is a ``quintessential'' scalar field presently rolling down
its potential \cite{quint}, for which particle physics models
have also been proposed \cite{mod1}.
The main motivation for constructing such dynamical schemes resides in
the hope of weakening the fine tuning issue implied by the
smallness of $\Lambda$.

In this respect, a very suitable class of models is provided by
inverse power scalar potentials, $V \sim \phi^{-q}$ 
with $q >0$, which admit attractor solutions in the scalar field
dynamics \cite{scalcosmo,track} characterized by a negative
equation of state. 
The behaviour of these solutions is determined by the cosmological
background and for this reason they have been denominated `trackers' in the
literature \cite{track}. 
A good feature of these models is that for a very wide range of the
initial conditions the scalar field will reach the tracking attractor
before the present epoch \cite{track}.
This fact, together with the negative equation of state, 
makes the trackers feasible candidates for explaining the
cosmological observation of a presently accelerating universe \cite{data}.
In this framework, indeed, it is possible to weaken the fine--tuning 
issue involved in choosing the initial conditions, even though 
the so--called `cosmic coincidence' problem still has to be fixed by hand. 
While the initial value of the scalar energy density is irrelevant if
the field comes on track before the present epoch, on the other hand
the point at which the scalar and matter energy densities are of the
same order depends on the mass scale in the potential. 
And this mass scale is fixed by requiring that 
$\Omega_\phi = {\cal O} (1)$ today.

It should be noted in passing that the requirement that the scalar
field has already joined the attractor {\it today} is a crucial
one. Indeed the field $\phi$ always passes though a `freezing' phase with 
$w_\phi = -1$ before eventually reaching the tracker \cite{track}.
Although $w_\phi = -1$ is compatible with the
present observational data, this value is indistinguishable from a
`true' cosmological constant and moreover it is not typical of the
trackers. Any scalar field sitting in a non-vanishing minimum of its
potential would give the same result, without any need of the dynamics
to bring it there.
In order to have the quintessence characteristic equation of state
$-1 < w_\phi < 0$, we should then require that tha scalar field $\phi$ is
already on track today.

As it will be discussed below, the range of initial conditions which 
allows $\rho_\phi$ to join the tracker before the present epoch 
is very wide.
Nevertheless, it should be noticed that in principle we do not have 
any mechanism to prevent $\rho_\phi^{in}$ from beeing outside the
desired interval. 
In this respect, an early universe mechanism which could
uniquely fix it at the end of inflation is needed.
In other words, if we find a way to naturally set $\rho_\phi^{in}$ 
in the range of values which allows for late time--tracking, we will
be assured that the `quintessence' field is a good candidate for the
unknown component which presently accelerates the universe.
\vskip.2in

A promising way to address the problem of initial conditions for
quintessence is the paradigm of ``quintessential inflation''
\cite{pv}, also referred to as the ``non oscillatory'' \cite{no} scheme.
The basic idea is to study an inflaton potential $V(\phi)$ which,
as it is typical in quintessence, goes to zero at infinity 
\cite{pv,no,qinfl}. 
In this way it is possible to obtain a late time quintessential
behaviour from the same scalar that in the early universe drives
inflation. The hope is that the end of inflation could uniquely fix
the initial conditions for the subsequent evolution of the scalar
componet of the universe.  

In ref. \cite{pv}, a model with a potential which goes like $\sim \phi^4$ for 
$\phi <0$ and like $\sim \phi^{-4}$ for $\phi >0$ is studied in
detail, and the authors use gravitational particle production 
for providing the entropy in the cosmological matter fields after 
the end of inflation. 
Although the shape of the potential for $\phi >0$ is the same studied
by Zatlev et al. in \cite{track}, they fail to 
show the `tracking' behaviour of the scalar field at late times 
because the initial conditions for the scalar energy density 
after inflation lay out of the phase space region that
leads to joining the attractor before the present epoch. 
Anyway they succeed in matching the present cosmological data 
because the cosmological constant-like behaviour ($w_\phi = -1$) 
that they find for the scalar field is also also a viable option. 
The reason why the scalar does not reach the tracker is the fact that
its energy density at the end of inflation is so low that it did not jet
have enough time to move towards the attractor.

The model in \cite{pv} suffers from some problems with respect to the
reheating mechanism that are extensively discussed in ref. \cite{no}. 
In particular the authors of \cite{no} propose 
to use the `instant preheating' \cite{inst} mechanism instead of 
gravitational particle production for the post-inflationary 
reheating phase. The main foucus of that work, though, is not on the
initial conditions for quintessence. 

In this paper we address the issue of the initial conditions for
quintessence in the context of the `quintessential inflation' paradigm.
The aim is two--fold. On one hand we will discuss under which
conditions an inflaton potential can leave a residual vacuum energy on
its tail, as already discussed in \cite{pv,no}. On the other we will
show that in some specific models it is possible to  have a late time
tracking ({\it i.e.} a well defined constrained behaviour of the scalar
and negative but $\not = -1$ equation of state) of the residual inflaton 
energy density.

After briefly recalling the constraints which inflation and
quintessence tracking models should separately meet, we will discuss to 
what extent they are compatible. 
We will then go on giving two specific examples, one in the context of
first order inflation an the other in the hybrid case.
In the first case we show that the `escape point' from the tunneling
naturally lies within the range which will produce a late time tracking. 
In the second, we re--examine the model proposed in
\cite{riotto} where a particle physics motivated potential was studied
in order to produce inflation. We find that in that model, a late time
quintessential behaviour is already built in and discuss under which
conditions it meets the observational constraints.

\subsection{Constructing workable models}

Before starting the construction of any unified model of inflation and 
quintessence, it is necessary to establish the constraints to which it 
should be subject. 
This is not a trivial task, since both inflation and quintessence model
building require very precise characteristics in order to be successful and
we must check if these separate needs are compatible 
with each other.
\vskip0.1in

Regarding {\it inflation}, there are four main points to be taken 
into account \cite{revinfl}:

\noindent
1. If we want the universe to be accelerating, the equation of state of 
the inflaton $\phi$
\begin{equation}
w_{\phi} = \frac{\dot{\phi}^{2}/2 -V(\phi)}{\dot{\phi}^{2}/2 +V(\phi)}
\end{equation}
must satisfy the inequality $w_{\phi} < -1/3$. This can be achieved if
$\dot{\phi}^2 < V(\phi)$.

\noindent
2. If inflation is to solve the flatness and horizon problems,
a sufficient number of e-foldings should take place. This means that 
the ratio of the final to the initial value of the scale factor $a$
must satisfy $a_{f}/a_{i} = \exp N$ with $N \gta 50$.

\noindent
3. The fact that the amplitude of scalar perturbations in the
cosmic microwave background, as measured by COBE in 1992, is of order 
$\sim 10^{-5}$ constrains the normalization of the inflaton 
potential.

\noindent
4. We must ensure that at the end of inflation 
sufficient reheating takes place. This is needed in order to produce 
the observed particle species in the universe. 
At the same time, one has also to check that gravitinos are not
overproduced. This puts on the reheating temperature an upper limit
which depends on the mass of the gravitinos and which is typically
around $10^{9} \div 10^{12}$ GeV \footnote{This bound refers to
thermal production. However, recently attention has been paid to gravitinos
production during preheating, suggesting that this mechanism could
overcome the thermal one \cite{grav}. Since this
 depends on the precise form of  the
superpotential of the whole supergravity theory, we will not deal with it in
the present work.}.

For what concerns {\it quintessence},  the following 
requirements should be fulfilled \cite{scalcosmo,track}:

\noindent
1. In order for the scalar field modeling of the cosmological constant
to be sufficiently general, we require that the post-inflationary shape 
of the potential is\footnote{This is the only class of potentials
that admits an analytic ``tracking'' attractor solution. 
However, more general cases have been studied in ref. \cite{track}.} 
\begin{equation}
V(\phi) \ \sim \ \frac{\Lambda^{4+q}}{\phi^{q}}\; , \;\;\;\;\;\; 
q > 0 
\; .  \label{q-potential}
\end{equation} 
In this way we are guaranteed that for a very wide range of initial conditions 
(indeed between the present critical energy density and the background 
energy density at the beginning, $\rho^{0}_{cr} \leq \rho^{in}_{\phi} 
\leq \rho^{in}_{\sc b}$) the scalar field will be rapidly
driven to a well-known ``tracking'' attractor behavior 
$\rho_{\sc tr}$:
\begin{equation} 
\frac{\rho_{\sc tr}}{\rho_{\sc b}} \ \propto \ 
a^{\frac{6(w_B +1)}{q+ 2}} \; ,
\label{scalelaw}
\end{equation}
where $w_B = 0, 1/3$ during MD and RD respectively. 
The attractor is characterized by an equation of state that during MD
is always negative: $w_{\sc tr} = (q\,  w_{\sc b} -2)/(q+2)$.
There are two main qualitative ways through which this can be
achieved (for more details see \cite{track}). 
If the initial conditions for $\phi$ are such that 
$\rho^{0}_{cr} \leq \rho^{in}_{\phi} \leq \rho^{in}_{\sc tr}$ 
({\it undershoot} case), then it will remain ``frozen'' until 
$\rho_{\phi} \sim \rho_{\sc tr}$ and then start to scale 
following eq. (\ref{scalelaw}). 
If, instead, initially $\rho^{in}_{\sc tr} \leq 
\rho^{in}_{\phi} \leq \rho^{in}_{\sc b}$ 
({\it overshoot} case) then $\phi$ will pass through 
a phase of kinetic energy domination before remaining frozen
at $\rho_{\phi} < \rho_{\sc tr}$ and eventually join the attractor.

\noindent
2. Secondly, we want the field $\phi$ to be already on track today
and its present energy density to correspond to what observations report, 
i.e. $\Omega_{\phi} \simeq 2/3$. 
These two conditions translate to 
\begin{equation}
V''(\phi) \simeq H^2 \;\;\;\;\;\;\; \mbox{and} \;\;\;\;\;\;\;  
V(\phi) \simeq \rho^0_{cr} \; , 
\label{q-conditions}
\end{equation}
which together imply for the quintessence field $\phi \simeq M_p$ today. 
Moreover, eq. (\ref{q-conditions}) provides a normalization for
the mass scale $\Lambda$ in the potential (\ref{q-potential}), giving
\begin{equation}
\Lambda \ \simeq \ \left( \rho^0_{cr} \ M_p^{\, q} \right)
^{\frac{1}{4+q}} \ \simeq\ 10^{-\frac{123}{4+q}}\ M_p
\; . \label{scalaL}
\end{equation}
This corresponds to choosing the desired tracker path to which the
scalar will be attracted to.
\vskip0.1in

While it is straightforward to find potentials with the required early  and
late-time behavior, the subtle issue resides in successfully matching  the exit
conditions for the scalar field after inflation with the range  of initial
conditions allowed for the trackers. For example, the naive guess of using the
potential $V = \Lambda^{4+ q} \phi^{-q}$ for quintessential
inflation is easily shown not to work. 
This is due to the fact that, with this potential,
the slow-roll conditions imply $\phi \gta M_p$ {\it during inflation}, while
the request that the quintessence field is presently dominating the universe
translates to $\phi \simeq M_p$ {\it today}.

The first scenario that we will discuss is first order inflation. 
In this context, if the potential $V(\phi)$ does not have an absolute
minimum but goes to zero as $\phi$ runs to infinity, the exit conditions
of the inflaton from the tunneling would set the starting
point for the subsequent quintessential evolution of the same field $\phi$.
If instead hybrid inflation is considered, it is again possible
to construct models in which the same field plays the role of the
inflaton and of the late time dynamical cosmological constant. 
In this case, the critical value $\phi_c$ that makes inflation stop
will determine the initial condition for the subsequent quintessential rolling.

\section{The models}
\subsection{First-order quintessential inflation}

In the original proposal of inflation \cite{guth}, the scalar
inflaton field $\phi$ which leads the expansion ``sits'' on a metastable
minimum of its potential $V(\phi)$ during the whole process. Inflation
eventually ends when bubbles of true vacuum nucleate through the
barrier and subsequently expand and collide reheating the universe.
A measure of the efficiency of the nucleation is given by
the ratio
\begin{equation}
\varepsilon = \frac{\Gamma}{H^4}
\end{equation}
between the tunneling rate $\Gamma$ and the Hubble constant
$H\,$.

As it was soon noticed \cite{exit}, models where $\varepsilon$ is
constant in time cannot work, because one needs both (i)
$\varepsilon \ll 1$ during inflation in order for the expansion
to last enough to solve the flatness and the horizon
problems and (ii) $\varepsilon \gta 1$ to have an efficient
nucleation. This puzzle is known as the ``graceful exit
problem''.
Many proposals have been suggested to solve this problem
(see \cite{kolb1} for a review), based on the possibility of
changing either $H$  \cite{extended} or $\Gamma$ \cite{double} with time.
This is commonly achieved by the use
of an ``auxiliary'' scalar field $\psi$, which is also employed to fit
the amplitude of scalar perturbations in the cosmic microwave background
measured by COBE\footnote{If this second field $\psi$ is slowly rolling 
down its own potential $V ( \psi)$, the amplitude of the
density fluctuations is given by (see \cite{muk} for a review)
$
10^{-\,5} \ \simeq\ \delta \rho /\rho \ \simeq \ 
\frac{3 \, H^3}{d V(\psi)/d\psi} \ \simeq \ 
3 \ ( 8 \pi /3 M_p^2 )^{3/2}  \: \frac{V
( \phi )^{3/2}}{dV(\psi)/d\psi} \;\;.
$
}.
Without entering the details of this procedure, we will fix $\varepsilon = 1$
in the toy model below, {\it assuming} that also in our case some auxiliary
field (or some other mechanism) can be invoked in this
regard\footnote{Since a late time quintessential behaviour obvoiusly
cannot affect the exit from inflation, the solutions proposed so far
(see for example \cite{extended,double}) may be assumed to work also 
in the present case.}.

In the model that we propose, the scalar field $\phi$ has a potential
(see fig. 1 below)
\begin{equation}
V \left( \phi \right) =
\frac{\Lambda^{\alpha+6}}{\phi^{\alpha} \left[ \left( \phi -
v \right)^2 + \beta^2 \right]} \;\; , \;\; \mbox{with} 
\;\;\frac{\beta}{v}\ll 1 \; ,
\label{firstpot}
\end{equation}
where $\Lambda$, $\beta$ and $v$ are constants of mass dimension one.
Eq. (\ref{firstpot}) has a barrier at $\phi \sim v$, after a
metastable minimum in $\phi \sim v\alpha /(\alpha +2) \equiv \phi_m$, 
while it behaves like $\sim \Lambda^{\alpha+6}/\phi^{\alpha+2}$ 
for $\phi \gg v$.

The parameter $\Lambda$ is constrained by quintessence (see eq. 
(\ref{q-conditions})-(\ref{scalaL})):
\begin{equation}
\Lambda \ \simeq \ 10^{-\frac{123}{\alpha+6}}\ M_p \ \simeq \ 
10^{\frac{19\:\alpha - 9}{\alpha +6}}\;\mbox{GeV} \;\;.
\end{equation}
In this way we ensure that the residual vacuum energy after inflation
does not overclose the universe and that at the same time it is not
presently negligibly small.

Inflation, instead, requires that most of the energy density 
$V(\phi_m)$ which dominates the accelerated expansion is transferred,
after the end of inflation, into a thermal bath of temperature $
T_{rh} \equiv 10^{9 +\gamma} \;\mbox{GeV}$ and this fixes the scale $v$ in the potential\footnote{For the degrees of freedom of the 
Standard Model $V \left( \phi_m \right) \simeq \rho_{rad} \simeq 35 \: 
T_{rh}^4$.} 
\begin{equation}
v \ \simeq \ ( \alpha + 2 ) \left( \frac{10^{-83
- 4 \, \gamma}}{140 \, \alpha^\alpha} \right)^{\frac{1} {\alpha+ 2}} M_p
\ \simeq \ ( \alpha + 2 ) \left( \frac{10^{19\alpha -45-4 \gamma}}{140
\ \alpha^\alpha} \right)^{\frac{1}{\alpha + 2}} \;\mbox{GeV} \;\;.
\label{v}
\end{equation}

The ratio $\beta / v$ is fixed below by the condition $\varepsilon =
1\;,$ so that the only free parameters of the model are the
exponent $\alpha$ and the reheating temperature parametrized
by $\gamma\,$.

Although the potential (\ref{firstpot}) does not have two minima (being
the lower one at infinity), the problem can be analytically approached in 
the so called ``thin wall limit'' \cite{coleman} as in the case in
which the two minima are present. This limit applies when the 
barrier is much higher than
the difference between the two minima. This is our case, since
\begin{equation}
\frac{V \left( v \right)}{V \left( \phi_m \right)} = \;\mbox{O} \left(
\frac{v}{\beta} \right)^2 \gg 1 \;.
\end{equation}
In order to get the decay rate $\Gamma$ (see \cite{coleman} for details)
one has to
integrate the equation of motion associated to the potential (\ref{firstpot})
and select the solution $\phi (x)$ which minimizes the Euclidean action
$S_E$ of the
system. This solution is O$\left(4 \right)$ symmetric (in the whole
Euclidean
space) and approaches $\phi = \phi_m$ at $x \equiv | {\bf x} | =
\infty\,$. The
value $\phi \left( x = 0 \right) \equiv \phi_{\rm es}$ is called the
``escape point'' and corresponds to the point at which the field $\phi$
tunnels out and starts rolling under its classical equation of motion.

If the thin-wall limit holds, the solution is 
$\phi \simeq \phi_{\rm es}$ for an interval $0 < x < R \,$, and then
$\phi \simeq \phi_m$ for $x > R \,$.
We physically interpret it as a bubble with radius $R$ of (nearly) true
vacuum within separated
by a thin wall from the false vacuum without. Continued to Minkowski space, 
the bubble appears to expand with a speed which asintotically approaches the 
speed of light. 
The universe can then be reheated by the particle production that occurs 
during the subsequent phase of collision of the bubbles recovering from 
the tunneling. 
The dynamics of this process in the present model is exactly analogous to the
one which occurs in the usual case (when the minimum of $V(\phi)$ is
at a finite value of $\phi$) and is extensively discussed in ref. 
\cite{bubbles}.

\begin{figure}[ht]
\begin{center}
\epsfig{file=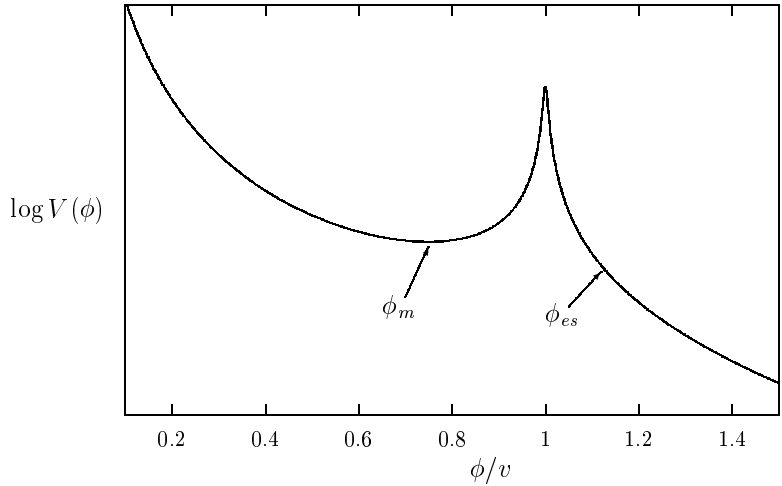,bbllx=115,bblly=480,bburx=490,bbury=715,height=7cm}
\begin{minipage}{14cm}
\begin{center}
\caption{\small\sl The bump in the potential of
eq. \ref{firstpot}, shown here with parameters $\alpha =6$ and 
$\beta /v =0.005$,
allows for an early stage of inflation while the inflaton field
$\phi$ sits in the relative minimum at $\phi_m$. After $\phi$ has
tunneled out at $\phi_{es}$, the quintessential phase starts with 
the scalar rolling down the slope $\sim \phi^{-\alpha -2}$ until today.}
\end{center}
\end{minipage}
\end{center}
\label{fig:pot}
\end{figure}

Following \cite{coleman}, the Euclidean action and the initial radius
of the bubbles are given by 
\begin{eqnarray}
S_E &\sim& - \, \frac{27 \, \pi^2}{2} \cdot \frac{S_1^4}{V(\phi_m)^3} 
\nonumber \\
R &\simeq& 3 \, S_1 / V ( \phi_m ) \;\;,
\end{eqnarray}
where
\begin{equation}
S_1 = \int_{\phi_m}^\infty d\,\phi 
\left[ 2 \left( V \left( \phi \right) - V \left( \phi_m
\right) \right) \right]^{1/2} \;\;.
\end{equation}

In our case $S_1$ can be calculated analytically for any value of
$\alpha$ in the potential (\ref{firstpot}) without any approximation, 
but a more readable and accurate enough estimate is given by
\begin{equation} \label{s1est}
S_1 \ \simeq\ 2 \int_{\phi_m}^v \,d\,\phi\,
\sqrt{2} \left[ \frac{\Lambda^{\alpha + 6}}{v^\alpha} \:
\frac{1}{\left( \phi - v \right)^2 + \beta^2} \right]^{1/2} \ =\
\frac{2\,\sqrt{2}\,\Lambda^{ \left( \alpha + 6 \right)
/2}}{v^{\alpha / 2}} \;\, \mbox{ln}\, \left[ \frac{4}{\alpha+2} \:
\frac{v}{\beta} \right] \;.
\end{equation}

The tunneling rate of $\phi$ is 
\begin{equation}
\Gamma \ =\ A \: e^{- \: S_E}
\label{tunrat}
\end{equation}
where $A$ is a parameter with mass dimension $4$ of order
$V( \phi_m )\,$.

From this equation, the condition 
\begin{equation}
\varepsilon \ =\ \frac{\Gamma}{H^4} 
\ =\ \left( \frac{3}{8\pi} \right)^2 \ \frac{M_p^4}{V(\phi_m)} \ e^{-S_E} 
= 1
\end{equation}
is obtained for $S_E \simeq 84 - 9 \: \gamma\,$, that is if the ratio
$\beta/ v$ satisfies
\begin{equation} \label{bv}
\mbox{ln}\: \left[ \frac{4}{\alpha+2} \: \frac{v}{\beta} \right] 
\ =\ \left( \frac{84 - 9 \, \gamma}{5.5 \cdot 10^5}\ 
140^{\frac{ \alpha +6}{\alpha + 2}} \right)^{1/4}  \left(
\frac{\alpha + 2}{\alpha^{\frac{\alpha}{\alpha + 2}}}
\right)^{\alpha /2}  \left( 10^{\alpha\,\left(\gamma-10\right)
+ 63 + 6\,\gamma} \right)^{1/(\alpha + 2)} \;.  
\end{equation}

Moreover, with this condition we also have
\begin{equation}
R \ =\ 0.36 \ \left( 84 - 9 \, \gamma \right)^{1/4} \,
10^{-\,9\,-\,\gamma}\ \mbox{GeV}^{-\,1}
\label{rad}
\end{equation}
as the analytical estimate for the initial radius of the bubbles\footnote{
In all this analysis we have not considered the
gravitational corrections on
the decay of the metastable vacuum. However, since they are of order $\left(
R H \right)^2\,$ \cite{coledelu} $\;$, their contribution is completely
negligible in all the cases of our interest.}.

Going on with the analysis, we specify to some particular values of the 
parameters.  As anticipated in Section $1.1$, we impose to the reheating 
temperature the  upper bound $T_{rh} \leq 10^{12}\, \mbox{GeV} \;$, 
that is $\gamma \leq 3$.
We see from eq. (\ref{bv}) that, for any fixed value of $\gamma$, it is
always possible to obtain
$\varepsilon = 1$ with arbitrarily low $\beta / v$, just 
allowing $\alpha$ to be large enough. 
However, for phenomenological reasons (see below) we  restrict ourselves to 
$\alpha \lta 10$ and list\footnote{In order to avoid excessive fine
tuning, we have not listed the cases for which $\beta/v<10^{-5}$. 
However this is a somewhat arbitrary limit and nothing
prevents from considering smaller values.} in Table 1 the cases for
which $\beta / v  \ll 1$.

\begin{table}[ht]
\begin{center}
\begin{minipage}{14cm}
\begin{center}
\begin{tabular}{|c|c|c|c|c|c|} \hline
$\gamma$ & $\alpha$ & $\beta /v$ & $R_{\mbox{an}} \left[ GeV^{-1} \right]$ &
$R_{\mbox{num}} \left[ GeV^{-1} \right]$ & $\phi_{es} / v$ \\
\hline
$1$ & $8$ & $7.75 \cdot 10^{-3}$ & $1.06 \cdot 10^{-10}$
& $0.98 \cdot 10^{-10}$ & $1.30$ \\
$2$ & $10$ & $9.72 \cdot 10^{-3}$ & $1.03 \cdot 10^{-11}$ & $0.94
\cdot 10^{-11}$ & $1.23$ \\
$3$ & $8$ & $4.74 \cdot 10^{-4}$ & $9.88 \cdot 10^{-13}$ & $9.07
\cdot 10^{-13}$ & $1.25$ \\ \hline
\end{tabular}
\caption{\small \sl
Comparison between the analytical and numerical results for the
radius of the tunneling bubble, for some values of the parameters
of the model. In the last column, the escape point is given in
units of $v$.} \label{tab1} 
\end{center}
\end{minipage}
\end{center}
\end{table}

We studied the tunneling also numerically and the solutions that we found are in good accordance with the previous semi-quan\-ti\-ta\-tive analysis. 
In particular, their shape is that of an instanton which interpolates between the initial value $\phi_{es}$ and the final one $\phi_m$.
The jump between the two values occurs at $x \equiv R_{num}$, very
close to the analytical estimate $R_{an}$  given by eq.~(\ref{rad}),
as can be checked in Table $1$.

At this stage, it is easily understood that the initial conditions for 
quintessence are entirely determined by the tunneling and not given 
arbitrarily. 
In particular, if the present model is considered, when $\phi$ tunnels
out at the escape point $\phi_{es}$ and starts rolling down the $V \sim
\phi^{-\alpha -2}$ potential, it has an energy density given by
\begin{equation}
V(\phi_{es}) \ \simeq\ V(\phi_m) \
\frac{4\, \alpha^\alpha}{(\alpha +2)^{\alpha +2}} \
\frac{(v/\phi_{es})^\alpha}{(\phi_{es}/v -1)^2}  \
\simeq\  \rho _{rad}\ \frac{4\,\alpha^\alpha}{(\alpha +2)^{\alpha +2}} 
\ \frac{(v/\phi_{es})^{\alpha+2}}{(1 - v / \phi_{es})^2}
\end{equation}
which is typically about $1 \%$ of the thermal one and which
can be computed as a function of $\alpha$ and of the reheating
temperature substituting the values of Table 1 in the last expression.
Note also that the field $\phi$ at the beginning of the
``quintessential'' regime is of the order of $v$ and, from eq. (\ref{v}),  
can be easily estimated to be $\ll M_p$ (actually it tends to
$M_p$ when $\alpha \rightarrow \infty$).

These initial conditions naturally lay within the allowed range for
quintessence and correspond to the {\it overshoot} case mentioned in
Section $1.1$. 
The field $\phi$ will then rapidly run to large values and its
energy density will consequently drop down well below the tracker, as
discussed in ref. \cite{track}. Then, after a ``freezing'' phase of
almost zero kinetic energy, it will eventually join the tracker path
at more recent times.
As a function of the exponent $\alpha$ in the potential (\ref{firstpot}), 
the equation of state of the scalar $\phi$ on the tracker is
\begin{equation}
w_\phi \ = \ - \frac{2}{\alpha + 4} \;\; .
\end{equation}
However it should be remembered that the present value of the equation
of state is lower than the attractor value. When the scalar energy
density ceases to be subdominant with respect to the matter one, the
approximation in which the attractor was derived does not hold
anymore \cite{track}. The scalar then leaves the tracking path as soon
as its energy density is comparable to that of matter and rapidly
tends towards a cosmological constant--like behaviour with $w_\phi =
-1$. For a present ratio $\Omega_\phi /\Omega_m \simeq 2$ we should
restrict to $\alpha \lta 10$ to be compatible with the present 
data \cite{data}.

\subsection{Hybrid quintessential inflation}

The model that we will consider next was proposed in \cite{riotto} and
involves a scalar potential arising from dynamical supersymmetry
breaking, of the form $V = V_{Susy}  + V_{\,\,\,\,\not\!\!\!\!Susy}$, 
with
\begin{equation}
V_{Susy} \ =\ M^4 \left| 1- \lambda \frac{\chi^2 \phi^2}{M^4} \right| ^2
\ +\ \frac{\Lambda^{4+p}}{\phi^p} \;\;\;\; , \;\;\;\;\;\;\;
V_{\,\,\,\,\not\!\!\!\!Susy} \ = \ \frac{1}{2} \beta M^2 \chi^2 \;\; .
\label{hybpot}
\end{equation}
As extensively discussed in \cite{riotto}, this potential can easily
accomodate an early inflationary stage of the  hybrid
\cite{hybrid} type\footnote{The presence of a SUSY--breaking mass 
term for the scalar $\phi$ is cosmologically excluded if we require 
a late time quintessential behaviour, since it would induce a minimum in
the $\phi$--direction of the potential.}.
We find that, quite surprisingly, this model has already incorporated
a late--time quintessential phase and this leads to important
consequences in addition to those discussed in \cite{riotto}. 
The interesting point is that this class of potentials is 
the first example which allows for discussing quintessential inflation
in a particle physics context.
While this issue is discussed both in the purely inflationary
(see the third reference in \cite{revinfl}) or purely quintessential 
\cite{mod1} cases, it is still missing in the `quintessential
inflation' scheme.
In the following we will then address this problem  with the potential
(\ref{hybpot}). 

For $\phi < \phi_c = \sqrt{\beta M^2/2\lambda}$ the minimum of the
potential is at $\chi =0$ and inflation can occur
if the term $M^4$ in (\ref{hybpot}) dominates.
When $\phi$ rolls down to $\phi > \phi_c$ inflation is ended by
instability in the $\chi$ direction, as typically occurs in hybrid models.
In this case, however, the VEV of the scalar potential $V$ does not 
almost instantaneously settle to zero but 
vanishes only after $\phi$ has run to infinity. 
This feature is very welcome if we want a quintessential component to be
present in the subsequent evolution of the universe.
In what follows we study for which range of the parameters this model
can fulfill the double aim of accounting for both the inflationary and
quintessential stages of our universe.

For $\chi =0$ and $\phi < \phi_c$, the potential can be rewritten as
\begin{equation}
V \ =\  M^4 \left[ 1 + \alpha \left( \frac{M_{p}}{\phi} \right)^p \right]
\label{vualpha}
\end{equation}
with 
\begin{equation}
\alpha \ \equiv\ \frac{\Lambda^{p+4}}{M_{p}^p \ M^4} \;\; . 
\label{alpha}
\end{equation}
We will see below that stringent upper limits apply to $\alpha$
for the model to fit observations. 
For the moment we only ask $\alpha$ to be small enough so that the
constant term dominates eq. (\ref{vualpha}), leading to a first inflationary
stage.
This is naturally achieved if 
we require that the term $\Lambda^{4+p}/\phi^p$ in eq. (\ref{hybpot})
leads to a present energy density that does not exceed the critical one
(see Section 1.1), that is if $\Lambda \leq \Lambda_c$ with
\begin{equation}
\Lambda_c^{p+4} \ =\ 10^{-\,123} \, M_p^{p+4} \;\;.
\end{equation}

To estimate the starting point of inflation we consider the slow roll 
parameters
\begin{eqnarray}
\varepsilon & \equiv & \frac{M_p^2}{4\,\pi}  \left( \frac{H^\prime 
( \phi )}{H ( \phi )} \right)^2
\simeq \frac{M_p^2}{16\,\pi}  \left( \frac{V^\prime 
( \phi )}{V ( \phi )} \right)^2  \ = \ 
\left( \frac{\phi_0}{\phi} \right)^{2 \, \left( p+1 \right)} \\
\eta & \equiv & \frac{M_p^2}{4\,\pi}  \left( \frac{H^{\prime\prime} 
( \phi )}{H ( \phi )} \right) 
\simeq - \frac{M_p^2}{8\,\pi}  \left[ \frac{V^{\prime\prime} 
( \phi )}{V ( \phi )} - \frac{1}{2} \left( \frac{V^\prime 
( \phi )}{V ( \phi )} \right)^2  \right]  \ =  \\ 
\nonumber
 &  & \ = \ \left( \frac{p+1}{2\sqrt{\pi}} \right)
\left( \frac{\phi_0}{\phi} \right)^{p+1} \frac{M_p}{\phi}
+ \left( \frac{\phi_0}{\phi} \right)^{2 \, \left( p+1 \right)}
\end{eqnarray}
where a prime denotes differentiation w.r.t. $\phi$ and
\begin{equation}
\phi_0 \ = \ \left( \frac{p}{4\,\sqrt{\pi}} \, \alpha \right)^{1/
(p+1)} \ M_p \;\;.  
\label{phio}
\end{equation}
Since slow roll requires $\varepsilon ,\eta \ll 1\;$, the accelerated 
expansion occurs for $\phi_0 \ll \phi \leq \phi_c \;$.  

When inflation ends at $\phi = \phi_c\;$, the second scalar $\chi$ 
leaves zero (that becomes an unstable maximum of V) and oscillates 
about one of the two new $\phi$--dependent minima that form at 
\begin{equation}
\chi^2_{MIN} \ =\ \frac{2 M^2}{\beta} \frac{\phi_c^2}{\phi^2} 
\left( 1 - \frac{\phi_c^2}{\phi^2} \right) \;\; .
\end{equation}
If we suppose that the scalar $\chi$ is coupled to the matter fields
via terms like $h\chi^2\varphi^2$, its coherent oscillations would
result in sufficient reheating of the universe by its decay products. 
The scalar $\chi$ eventually settles to $\chi_{MIN}$ and the reheating 
temperature is typically of order $\sim M$.
After the reheating phase is completed the potential is given by
\begin{equation}
V(\chi_{MIN}, \phi) \ =\ M^4 \frac{\phi_c^2}{\phi^2}  
\ + \ \frac{\Lambda^{4+p}}{\phi^p} \;\; .  
\label{potmin}
\end{equation}
It should be noted that, taking into account the bound (\ref{reslam}) 
on $\Lambda$ given below, it turns out that at late times ($\phi
\gg \phi_c$) the term that will dominate in the above formula is the
first one. That means that the tracking behavoiur will not be induced,
as might be expected, by the explicit inverse power term present in the
potential, but will instead come from the effective contribution $\sim
\phi_c^2/\phi^2$ originated by the minimum of the second scalar $\chi$.

The first term in eq. (\ref{potmin}) constrains the ratio of $\phi_c\,$
to $M$, since we need that today $\;V(\chi_{MIN}, \phi \simeq M_p) \simeq
\rho^{0}_{cr}\;$. This translates to
\begin{equation}
\frac{\phi_c}{M} \ =\ \sqrt{\frac{\beta}{2\,\lambda}} \ \simeq\
 \left( \frac{100 \:\mbox{GeV}}{M} \right)^3 \cdot 10^{-10} \;\;.
\label{rescoef}
\end{equation}
The smallness of $\phi_c$ constrains the other parameters of the model, once
the two main requirements of inflation, namely that it lasts for enough
e--folds and that it gives the spectrum of fluctuations observed by COBE, 
are taken into account. 
The number of e--folds of inflation, when $\eta \gg \varepsilon$ is given by
\cite{riotto}
\begin{equation}
N_{\mbox{tot}} \ =\ \frac{2 \, \sqrt{\pi}}{M_p} \: 
\int_{\phi_{in}}^{\phi_c} \frac{ d \, \phi^\prime}{ \sqrt{\varepsilon 
\left( \phi^\prime \right)}} \ \simeq\
\frac{8\pi}{p(p+2)} \frac{1}{\alpha} \left( \frac{\phi_c}{M_p} \right)^{p+2} 
\;\;. \label{ntot}
\end{equation}
Since at least  $50$ or $60$ e--folds are needed to solve the horizon and 
flatness problems, we impose $N_{\mbox{tot}} \geq 50\;$. 
For what concerns the fluctuations, instead, we require that the 
curvature power spectrum\footnote{$\phi_{50}$
is the value of the inflaton field $50$ e--folds before the end of inflation.}
\cite{riotto}
\begin{equation}
P_R^{1/2} \equiv \frac{1}{\sqrt{\pi}} \, \frac{H \left( \phi_{50} \right)}
{M_p \, \sqrt{\varepsilon \left( \phi_{50} \right)}}\ \simeq\ \frac{p+2}{\pi} 
\sqrt{\frac{2 \, \pi}{3}} \, \left( \frac{M^2}{M_p \, \phi_c} \right) \, 
N_{\mbox{tot}} \, \left( 1 - \frac{50}{N_{\mbox{tot}}} \right)^{(p+1)/(p+2)}
\label{cobe}
\end{equation}
matches the COBE normalization \cite{cobe} $P_R^{1/2} = 5 \cdot
10^{-\,5}\;$, with spectral index \cite{riotto}
\begin{equation}
n - 1 \ \equiv \ \frac{d\ \log (P_R)}{d\ \log k} \ = \
- 4 \varepsilon + 2 \eta \ \simeq \ \left( \frac{p+1}{p+2} \right) \
\frac{2}{N_{\mbox{tot}} (1- 50/N_{\mbox{tot}})}  \;\; .
\end{equation}
The spectrum turns out to be blue ($n > 1$), but for $N_{\mbox{tot}}
> 50$ it quickly approaches scale invariance, $n \simeq 1$.
The present limit $|n-1| <0.2 $ \cite{revinfl} translates to
$N_{\mbox{tot}} \gta 60$.  

Substituting eqs. (\ref{alpha}) and (\ref{rescoef}) into eq. (\ref{ntot}),
we obtain the following upper bound on the mass scale\footnote{Inserting
this value for $\Lambda$ in eq. (\ref{phio}) we can check that,
consistently, $\phi_0 \ll \phi_c\;$.}  $\Lambda$:
\begin{equation}
\left( \frac{\Lambda}{\Lambda_c} \right)^{p+4} 
\ \simeq \ \frac{10^{-27 p}}{p (p + 2) 2^{p+1}} 
 \left( \frac{100 \, \mbox{GeV}}{M} \right)^{2p} \, 
\left( \frac{50}{N_{\mbox{tot}}} \right)
\;\;.
\label{reslam}
\end{equation}
Finally, from eq. (\ref{cobe}) we get  the rough estimate
$\frac{M^2}{M_p \, \phi_c} \: N_{\mbox{tot}} \sim 10^{-\,5}\,$, which
translates into
\begin{equation}
M \ \simeq\ 100 \:\mbox{GeV} \ \left( \frac{50}{N_{\mbox{tot}}}  \right)^{1/4} \;\;.
\end{equation}
We thus understand that a sufficient amount of e--folds can be achieved 
only for a quite low reheating temperature (remember $T_{rh} \simeq M$). 
Anyhow, a low $T_{rh}$ is also preferred since it weakens the upper bounds 
on $\phi_c$ and on $\Lambda$ given 
by eqs. (\ref{rescoef}) and (\ref{reslam}).

As an example of the orders of magnitude involved, for $M \simeq 50
\:\mbox{GeV}\,$ we get $N_{\mbox{tot}} \simeq 800\:$, $\; \phi_c
\simeq 22\:\mbox{eV}\:$, and $\Lambda$ in the range $0.016 \:\mbox{eV} 
(p = 2)$ $\div \: 5 \:\mbox{eV} (p = 50)$. The last mass scale is not very
natural in a supersymmetric context, where one customarily expects
values $\gta \mbox{GeV}$. 
However, the amount of fine tuning involved in $\Lambda$ from eq. 
(\ref{reslam}) is milder than in the case of the cosmological 
constant, where the mass scale is more than
30 orders of magnitude smaller than the ``natural'' value.

The last important point in our discussion is the ``quintessential''
evolution of $\phi$ after the reheating phase. 
As already noticed, the bound on $\Lambda$ given by eq. (\ref{reslam})
forces the second term in eq. (\ref{potmin}) to be completely
negligible during this last phase. 
Despite its shape is exactly the one required for the trackers, 
the only role it plays in this model is to drive $\phi$ towards $\phi_c$
during inflation. 
The term which dominates the potential (\ref{potmin}) at late times 
comes instead from the dynamics of the field $\chi$ and the tracking 
behavior is guaranteed from the fact that it involves a negative power
of the inflaton $\phi\,$ as well.

The initial conditions of this ``quintessential'' phase are fixed 
by the value $\phi^{*}$ of the field $\phi$ after reheating, when $\chi =
\chi_{MIN}$ and eq. (\ref{potmin}) starts holding. 
The precise value of $\phi^{*}$ depends on the details of the physics 
which governs the reheating, but it is reasonable to assume that it will
not be much larger than $\phi_c\,$. 
If this is the case, the initial energy of the quintessential field $\phi$
will be somewhat smaller (but not too smaller) than the one stored in
the thermal background and, as in the previous model, we are again in the
``overshoot'' case.

The attractor equation of state for a potential $V \sim \phi^{-2}$ is
simply $w_\phi = -1/2$, well within the observational bound \cite{data}.

\section{Conclusions}

In this paper we have discussed two possible schemes in which inflation
and quintessence are unified. In both cases it is the same field which 
at the same time plays the role of the inflaton and of the quintessence 
scalar. In this way we succeded to uniquely fix the initial conditions for
quintessence from the end of inflation and have found that they are
compatible with a late--time tracking.

In one example we studied first-order inflation with a potential going  to
zero at infinity like $\phi^{-\alpha}$. A bump in the potential at  $\phi
\ll M_p$ allows for an early stage of inflation while the scalar  field
gets ``hung up'' in the metastable vacuum of the theory.  Nucleation of
bubbles of true vacuum through the potential barrier  sets the end of
the accelerated expansion and starts the reheating phase.  As it is well
known, this scenario suffers from the so-called ``graceful exit problem'', 
but we briefly commented on possible ways out where (thanks to some
auxiliary scalar field) the  ratio of the tunneling rate to the Hubble
volume, $\Gamma/H^4$, varies with time. 
After the reheating process is completed, the
quintessential  rolling of the scalar $\phi$ starts and its initial
conditions (uniquely  fixed by the end of inflation) are 
naturally within the range which leads to a tracking behavior in recent times.

As an alternative, we considered the model of hybrid inflation which,
motivated by dynamical supersymmetry breaking, was proposed by the authors
of \cite{riotto}. We showed that it naturally includes a 
late--time quintessential behavior. 
This result is very interesting since it is the first time that  the 
quintessential inflation scheme is discussed in a particle 
physics motivated context. 
As typical of hybrid  schemes, the potential is dominated 
at early times (that is until the inflaton field is smaller than a 
critical value $\phi_c\,$) by a constant term and inflation
takes place. 
Eventually the inflaton rolls above $\phi_c\,$, rendering unstable the
second scalar of the model, $\chi$. 
This field starts oscillating about its minimum (whose position is
determined by $\phi$) and in this stage the universe is reheated. 
After $\chi$ has settled to the minimum, the inflaton continues its slow
roll down the ``residual'' potential which goes to zero at infinity like
$\phi^{-2}$, thus allowing for a quintessential tracking solution. 
Also in this case the initial conditions for the quintessential 
part of the model do not have to be set by hand, but depend uniquely on
the value of the inflaton field at the end of reheating.

In our analysis we have shown that in both models inflation and
quintessence can be accounted for within a relatively wide range of the
parameters of the potential.
Some differences exist though. 
The first one concerns the exponent $p$ characterizing
the late time shape of the potential $V (\phi) \sim \phi^{-p}$  
which leads the quintessential part of the models.  
While in the first case the exponent can be read directly from the 
potential, in the other the leading contribution comes from the 
dynamics of the second scalar $\chi$ which gives $V \sim \phi^{-2}$.
The explicit term proportional to $\phi^{-p}$ in the potential of this last
model  is just needed for driving the inflaton towards $\phi_c$ during
inflation and then is completly negligible. 
Moreover, contrary to the first scheme, the dependence of the physical 
quantities on $p$ is in this case very mild and hardly 
distinguishable by observations.

Another important difference between the two analyses is that in the
first-order model we allowed for relatively high 
reheating temperatures, while in the second case we obtained
$T_{rh} \lta 100 \,\mbox{GeV}\,$. A low reheating temperature
can still be accommodated in a successful cosmology, since the only model
independent limit is $T_{rh} \gta \mbox{GeV}$ in order to produce protons
and neutrons. Moreover such  small values could avoid the restoration of
symmetries which in many extensions of the Standard Model 
would produce a number of unwanted topological defects.

\vskip0.4in
The authors are grateful to Antonio Riotto for a number of helpful
discussions and for carefully reading the manuscript. We also thank
Daniele Montanino for some technical help in the numerical part of this
work.


\begin{thebibliography}{99}

%
%%%%%%%%dati cosmo
%
\bibitem{obs}
 N.A.~Bahcall, J.P.~Ostriker, S.~Perlmutter and P.J.~Steinhardt, 
 Science {\bf 284}, 1481 (1999).

%%%%%%%quintessence
%
\bibitem{quint} 
 J.A.~Frieman and I.~Waga, Phys.  Rev. {\bf D57}, 4642 (1998); 
 R.R.~Caldwell, R.~Dave, and P.J.~Steinhardt, Phys. Rev. Lett. {\bf 80}, 
 1582 (1998); 
 M.S.~Turner and M.~White, Phys. Rev. {\bf D56}, 4439 (1997);
 F.~Perrotta, C.~Baccigalupi and S.~Matarrese, {\sl astro-ph/9906066};
 R.~de Ritis, A.A.~Marino, C.~Rubano, P.~Scudellaro {\sl
 hep-th/9907198};
 T.~Chiba, N.~Sugiyama and T.~Nakamura, Mon. Not. Roy. Astron. Soc. 
 {\bf 289} (1997) L5.

%%%%%%%%%%%modelli
%
\bibitem{mod1} 
 P.~Bin\'{e}truy, {\sl hep-ph/9810553}; 
 A.~Masiero, M.~Pietroni and F.~Rosati, {\sl hep-ph/9905346};
 P.~Brax and J.~Martin, {\sl astro-ph/9905040}; 
 J.A.~Frieman, C.T.~Hill, A.~Stebbins, and I.~Waga, Phys. Rev. Lett. 
 {\bf 75}, 2077 (1995); 
 K.~Choi,  {\sl hep-ph/9902292};  
 J.E.~Kim, JHEP {\bf 9905}, 022 (1999); 
 J.E.~Kim, {\sl hep-ph/9907528};
 M.C.~Bento and O.~Bertolami {\sl gr-qc/9905075};
 V.~Sahni and S.~Habib, Phys. Rev. Lett. {\bf 81} (1998) 1766.

%%%%%%%scalar cosmo
%
\bibitem{scalcosmo} 
 P.J.E.~Peebles and  B.~Ratra, Astrophys. Jour. {\bf 325}, L17 (1988); 
 B.~Ratra and P.J.E.~Peebles, Phys. Rev. {\bf D37}, 3406 (1988);
 A.R.~Liddle and R.J.~Scherrer, Phys. Rev. {\bf D59}, 023509 (1999).

%%%%%%%%trackers
%
\bibitem{track} 
 I.~Zlatev, L.~Wang, and P.J.~Steinhardt, Phys. Rev. Lett. {\bf82}, 896 (1999);
 P.J.~Steinhardt, L.~Wang, and I.~Zlatev, Phys. Rev. {\bf D59}, 123504 (1999). 

\bibitem{data}
 L.~Wang, R.R.~Caldwell, J.P.~Ostriker and P.J.~Steinhardt,
 astro-ph/9901388.

%%%%%%%%quint-inflation
%
\bibitem{pv}
 P.J.E.~Peebles and A.~Vilenkin, Phys. Rev. {\bf D59}, 063505 (1999).
\bibitem{no} 
 G.~Felder, L.~Kofman and A.~Linde, {\sl hep-ph/9903350}.
\bibitem{qinfl}
 L.H.~Ford, Phys. Rev. {\bf D35}, 2955(1987); 
 B.~Spokoiny, Phys. Lett. {\bf B315}, 40 (1993); 
 M.~Joyce and T.~Prokopec, Phys. Rev. {\bf D57}, 6022 (1998).

%%%%%%%%%%%preheating
%
\bibitem{inst} 
 G.~Felder, L.A.~Kofman and A.D.~Linde, Phys. Rev. {\bf D59}, 123523 (1999). 

%
%
\bibitem{riotto}
 W.H.~Kinney and A.~Riotto, Asrotpart. Phys. {\bf 10}, 387 (1999);
 W.H.~Kinney and A.~Riotto, Phys. Lett. {\bf B435}, 272  (1998).

%
\bibitem{revinfl}
 For a review see for example: 
 E.W.~Kolb, {\sl astro-ph/9612138};
 A.R.~Liddle, {\sl astro-ph/9901124};
 D.H.~Lyth, A.~Riotto, Phys. Rept. {\bf 314}, 1 (1999).

%
\bibitem{grav}
 R.~Kallosh, L.A.~Kofman, A.D.~Linde, A.~Van Proeyen, {\sl
 hep-th/9907124};
 G.F.~Giudice, I.~Tkachev and A.~Riotto {\sl hep-ph/9907510}.

%%%%%%%%%1st order inflation
%
%
\bibitem{guth}
 A.~Guth, Phys. Rev. {\bf D23}, 347 (1981). 
\bibitem{exit}
 S.W.~Hawking, I.G.~Moss, J.M.~Stewart, Phys. Rev. {\bf D26}, 2681 (1982);
 A.H.~Guth, E.J.~Weinberg, Nucl. Phys. {\bf B212},  321 (1983). 
\bibitem{kolb1}
 E.W.~Kolb, Physica Scripta {\bf T36}, 199 (1991).
\bibitem{extended}
 D.~La and P.J.~Steinhardt, Phys. Rev. Lett. {\bf 72}, 376 (1989).
\bibitem{double}
 A.D.~Linde Phys. Lett. {\bf B249}, 18 (1990);
 F.~Adams and K.~Freese, Phys.Rev. {\bf D43}, 353 (1991). 
\bibitem{muk}
 V.F.~Mukhanov, H.A.~Feldman and R.H.~Brandenberger,  Phys. Rept. 
 {\bf 215},  203 (1992).
\bibitem{coleman}
 S.~Coleman, Phys. Rev. {\bf D15}, 2929 (1977); 
\bibitem{bubbles}
 R.~Watkins and L.M.~Widrow, Nucl. Phys. {\bf B374}, 446 (1992);
 M.S.~Turner, E.J.~Weinberg and L.M.~Widrow, Phys. Rev. {\bf D46}, 2384
 (1992);
 A.~Masiero and A.~Riotto, Phys. Lett. {\bf B289}, 73 (1992);
 E.W.~Kolb, A.~Riotto and I.~Tkachev,  Phys. Rev. {\bf D56}, 6133 (1997).
\bibitem{coledelu}
 S.~Coleman and F.~De Luccia Phys. Rev. {\bf D21}, 3305 (1980). 

%%%%%%%%%hybrid
%
%\bibitem{riotto}
% W.H.~Kinney and A.~Riotto, Asrotpart. Phys. {\bf 10}, 387 (1999);
% W.H.~Kinney and A.~Riotto, Phys. Lett. {\bf B435}, 272  (1998).
\bibitem{hybrid}
 A.~Linde, Phys. Lett. {\bf B259}, 38 (1991); 
 A.~Linde, Phys. Rev. {\bf D49}, 748 (1994).
\bibitem{cobe}
 E.F.~Bunn, M.~White, Astrophys. Jour. {\bf 480}, 6 (1997). 

\end{thebibliography}
\end{document}